\documentclass[aps,prb,twocolumn,
	groupedaddress,superscriptaddress,
	amsfonts,amssymb,amsmath,floatfix,
	citeautoscript]{revtex4-1}
\usepackage{graphicx}
\usepackage[centering,hmargin=19mm,tmargin=28mm,bmargin=28mm]{geometry}
\usepackage{mathtools}
\usepackage{braket}
\usepackage{float}
\usepackage{hyperref}
\usepackage{color}
\usepackage[normalem]{ulem}
\usepackage{multirow}
\usepackage{newtxtext}
\usepackage{caption}
\usepackage{subcaption}
\usepackage{booktabs}
\usepackage{xcolor}
\usepackage{amsmath}
\usepackage[normalem]{ulem}
\usepackage[textwidth=1.5cm,textsize=footnotesize]{todonotes}
\hypersetup{colorlinks, 
    linkcolor={blue!75!black!80!yellow},
    citecolor={blue!75!black!80!yellow}, 
    urlcolor={blue!75!black!80!yellow}
    }
\usepackage[cmintegrals]{newtxmath}

\makeatletter
\renewcommand\@make@capt@title[2]{%
    \@ifx@empty\float@link{\@firstofone}{\expandafter\href\expandafter{\float@link}}%
    \sffamily{\textbf{#1}}\@caption@fignum@sep#2
}%

\makeatother

\newcommand{\HarvardSEAS}{John A.
Paulson School of Engineering and Applied
Sciences, Harvard University, Cambridge, MA, USA}

\begin{document}

\author{Michael Cubeddu}
\thanks{These two authors contributed equally}
\email{mcubeddu@college.harvard.edu}

\author{Will Finigan}
\thanks{These two authors contributed equally}
\email{finigan@college.harvard.edu}

\author{Thomas Lively}
\altaffiliation{Present address: Google LLC, 1600 Amphitheatre Parkway, Mountain View, CA 94043}
\affiliation{\HarvardSEAS}

\author{Johannes Flick}
\email{flick@seas.harvard.edu}
\affiliation{\HarvardSEAS}

\author{Prineha Narang}
\email{prineha@seas.harvard.edu}
\affiliation{\HarvardSEAS}

\title{Introducing Control Flow in Qubit Allocation for Quantum Turing Machines}

\date{\today}

\begin{abstract}
Different platforms for quantum computation are currently being developed with a steadily increasing number of physical qubits. To make today's devices practical for quantum software engineers, novel programming tools with maximal flexibility have to be developed. One example to extend the applicability of quantum computers to more complex computational problems is quantum control flow. The concept of control flow allows for expanded algorithmic power of the programming language in the form of conditional statements and loops, which a linearly-executed program is incapable of computing. In this work, we introduce a framework to reconcile the non-deterministic properties of quantum control flow when allocating logical qubits from a given quantum circuit to a specific NISQ device in the pre-processing and compiling stage. We consider the respective connectivity and fidelity constraints, with the goal of reducing the expected error rate of the computation. This work will allow for quantum developers and NISQ devices together to more efficiently exploit the compelling algorithmic power that the \textit{quantum Turing machine} model provides.
\end{abstract}

\maketitle
\section{Introduction}
\label{sec:intro}
In recent years, the field of quantum information science has gathered rapid momentum and now allows the realization of quantum algorithms on various quantum hardware. Different platforms for quantum computation are currently being developed, with the leading platforms in scalability being superconducting systems~\cite{kandala2017,zeng2017,hempel2018}, ultra-cold atoms~\cite{cotler2018}, photonic systems~\cite{pichler2017,carolan2019} and trapped ions~\cite{nam2019}. These systems are well-positioned to simulate systems that cannot currently be computed on classical computers, with a search for the elusive quantum advantage underway~\cite{dalzell2018, reiher2017}. Simultaneously, quantum algorithms have been proposed for a diverse set of applications, ranging from chemistry~\cite{cao2018}, material science~\cite{babbush2018}, machine learning~\cite{schuld2018}, algebra~\cite{harrow2009}, to the quantum internet~\cite{dahlberg2019}. While the advent of digital quantum computers promises revolutionary new possibilities, the current state of the hardware can be best described as ``noisy-intermediate scale quantum'' (NISQ)~\cite{preskill2018} devices. Therefore, an important direction for the field is to develop algorithms for such NISQ devices, and to explore strategies to most efficiently utilize current physical resources for problems of interest.

In general, quantum algorithms consist of sets of quantum gates and circuits~\cite{deutsch1985, selinger2004}.
A successive, linearly executed set of quantum gates and circuits focuses on the flow of the data, but neglects the control flow. Control flow is the order in which individual instruction blocks are evaluated. As a result of the evaluation of a control flow statement, the algorithm makes a choice which of the possible path are evaluated next. Control flow integration allows for expanded algorithmic power of the programming language in the form of conditional statements and loops; features that a linearly-executed basic block is incapable of computing. This integration expands the computational functionality of the program much further than a simple basic block. Additionally, control flow is necessary to make a programming language Turing-complete, that is to make it able to simulate a Turing machine~\cite{turing1936}. With quantum control flow, these conditional statements may depend on the measurements of qubits, adding a probabilistic effect to the program that one may take advantage of to solve more complex problems or to solve problems faster than its classical counterpart. Because the native gate set of quantum Turing machines subsumes the native gate set for classical Turing machines, moving towards the quantum model offers the potential to unlock more computational and algorithmic power.

The noise in the quantum computation on current NISQ devices, both systematic and stochastic, leads to high error rates in the quantum computation and is due to a combination of factors including decoherence, dephasing of the qubit states, calibration errors, and readout and gate infidelities. The presence of both systematic and stochastic sources of noise introduces drastic limits on the number of actions/operations that can be performed on qubits before the result becomes noise. As a consequence, these noise constraints of NISQ devices necessitate theoretical schemes that make most efficient use of the computational resources by minimizing the error rates to enhance the probability of successful algorithm execution.

A promising possibility to effectively reduce error rates is to find an efficient allocation scheme that maps logical qubits from a given quantum circuit to the physical qubits on a specified quantum device. The so-called \textit{qubit allocation problem} is the problem of minimizing the total error rate of a given quantum program considering the connectivity constraints and different error rates of the individual qubits for a specific quantum device. Finding optimal solutions to several variants of the \textit{qubit allocation problem} requires solving multiple NP-hard sub-problems~\cite{siraichi2018}. Further, limited connectivity between the physical qubits in NISQ devices necessitates the insertion of SWAP gates, further complicating the problem~\cite{siraichi2018}. Therefore, different research groups have recently proposed different allocation schemes~\cite{zulehner2018, siraichi2018,shafaei2013, li2018, finigan2018, guerreschi2018, murali2019, murali2019b, li2018}, although none have considered quantum control flow.

To address this critical gap in the field of quantum information science, in this \emph{Article} we introduce a first practical scheme to optimally allocate quantum programs on NISQ devices that include quantum control flow. As quantum computers become more and more reliable and powerful, quantum software engineers will be able to run increasingly complex programs on quantum computers. Consequently, it is important that these software developers have the most expressive and efficient Turing-complete languages at their disposal.

The layout of this \emph{Article} is as follows: We first introduce the control flow graph processing, its connection to the qubit allocation problem and implications on the SWAP insertion. In the second part of this paper, we benchmark our scheme.

\begin{figure}[ht]
      \includegraphics[width=0.45\textwidth]{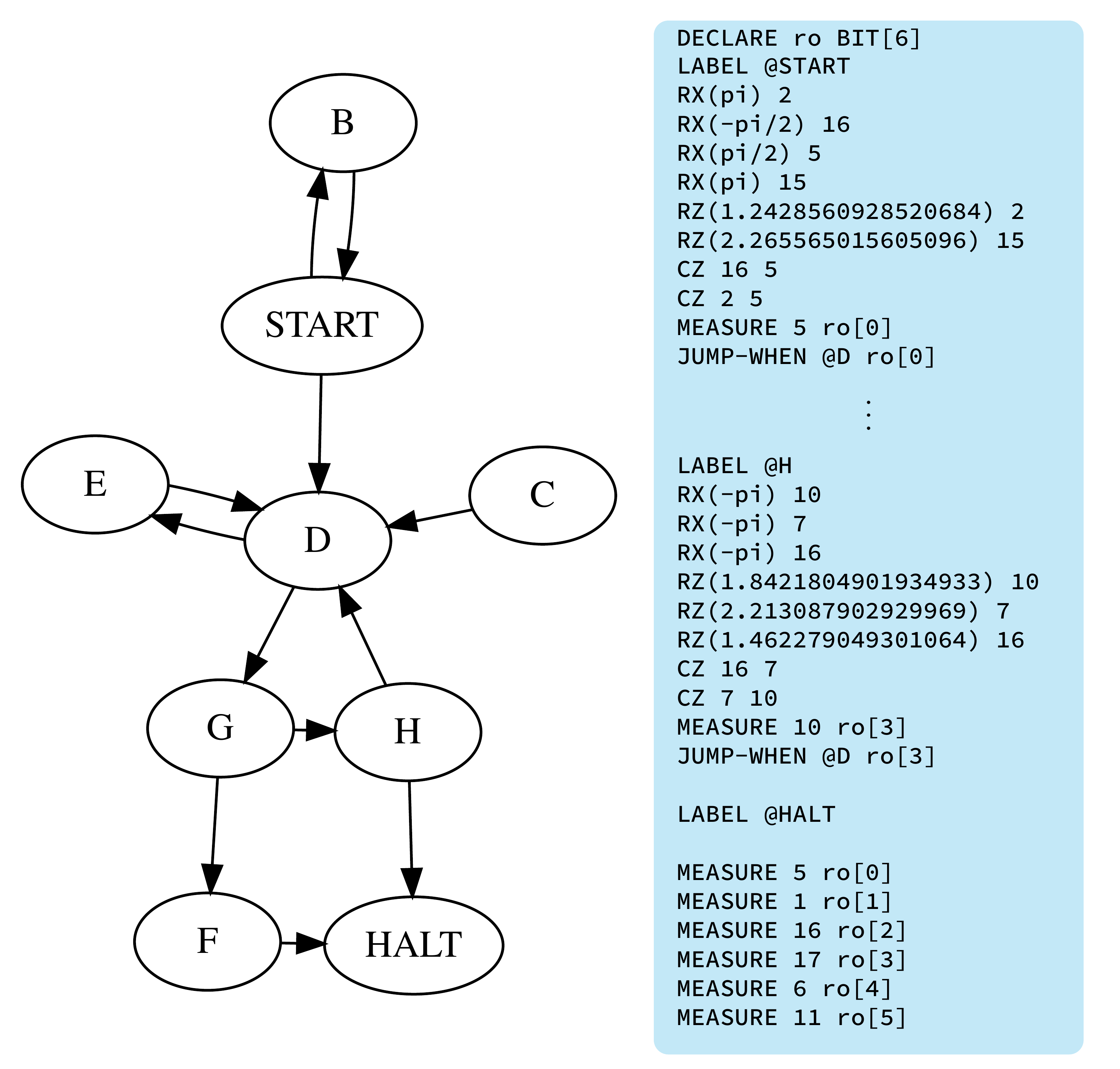}
      \caption{
      Left: Control flow graph for generated test circuit from START to HALT. Each basic block is assigned with a letter. Right: Description of generated test circuit in \texttt{Quil} instruction language.}
      \label{fig:control-flow}
 \end{figure}
\section{Qubit allocation and Control Flow}
\label{sec:cfgp}

An example of a quantum programming language that is currently commercially available and capable of simulating quantum control flow is the Quil language~\cite{smith2017} and in the following, we will use Quil to exemplify our strategy for a first practical quantum control flow scheme. There also exist several other quantum programming languages that include support for control flow~\cite{selinger2004, fu2018} and we the reader to Ref.~\cite{larose2018} for an overview of different quantum software platforms and their current capabilities.

In Fig.~\ref{fig:control-flow}, we illustrate a quantum circuit that includes quantum control flow on the left. The program will run from \textsc{START}, to \textsc{HALT}, and consists of connected blocks. The specific path that the program is executes depends on the outcome of the measurements that are performed during the execution of these blocks. The Quil language provides the instructions \textsc{JUMP}, \textsc{JUMP-WHEN}, and \textsc{JUMP-UNLESS} to realize control flow in the program. In Fig.~\ref{fig:control-flow} on the right, we show an example of a quantum program written in Quil including quantum control flow. The locations in the program that these instructions can jump to are identified with specific \textsc{LABEL}s.
The two-conditional jump instructions \textsc{JUMP-WHEN} and \textsc{JUMP-UNLESS} either direct the flow of the program to a given label, or continue the flow to the next instruction, depending on the result of a measurement of a qubit.

These control flow primitives are sufficient to implement loops and branching constructs of higher level programming languages and make the Quil language Turing-complete, and at least as powerful as the classical Turing machine, since the set of classical logic operators is a subset of the available operations in the Quil Language.

\subsection{Allocating Control Flow}
% \label{sec:bbweights}

In a previous work~\cite{finigan2018}, we introduced an efficient allocation scheme using the simulated annealing method with local search of the solution space using Dijkstra's algorithm. The algorithm takes into account the weighted connectivity constraints of both the quantum hardware and the quantum program, and considering the swap-placement with the lazy swap insertion strategy. This scheme takes advantage of the construction of the coupling graph, where all edges are associated with an error rate (fidelity) due to the different qubit quality.

Control flow introduces the following challenge to the qubit allocation problem: In linear programs every instruction in the program is executed exactly. However if control flow is present in the program this assumption is not valid anymore. For quantum programs, depending on the results of non-deterministic measurements of qubits during execution of the program, a program may execute instructions arbitrarily many times, including zero. This is a fundamental challenge for any allocation scheme since it must take into account the relative influence of qubit connectivity constraints.

For example, if a program contains ten two-qubit gates operating on qubits $A$ and $B$ and only one gate operating on qubits $C$ and $D$, the naive approach would assume that qubits $A$ and $B$ should be allocated to the physical qubit pair with the higher fidelity. However, that is not the case if the gate operating on qubits $C$ and $D$ is executed a hundred times more than any other gate due to the control flow. Since the exact number of executions of each gate cannot be determined prior to the execution of the program due to the probabilistic nature of qubits, any qubit allocation scheme must consider the expected number of executions of each instruction.

To further elucidate how to calculate the expected number of executions, we will first define some useful concepts from the literature on compilers~\cite{allen1970, allen1971, aho2008}.
A \textit{basic block} is a contiguous sequence of instructions that can only be entered at its beginning and is exited at its end.
Since there are no jumps into or out of the middle of a basic block, all instructions in a basic block are executed the same number of times.
A basic block $b$ has a set of predecessors, $\mathrm{pred}(b)$, and successors, $\mathrm{succ}(b)$.
The predecessors of $b$ are the basic blocks that jump to $b$ and the successors of $b$ are those blocks that $b$ can jump to.
Together, the set of basic blocks in a program are the vertices in a \textit{control flow graph (CFG)}, a directed graph that describes how control can move through the program, as illustrated in Fig.~\ref{fig:control-flow} on the left.

Using these definitions we need to find the expected number of executions of each basic block in a single execution for an effective qubit allocation scheme.
Let $F_b$ be the expected number of executions of a basic block $b$ and $P_{b,b^\prime}$ be the probability that control transfers from block $b$ to block $b^\prime$. Then, the expected number of executions of a basic block $b$ is given by

\begin{align}
F_b = \sum_{p \in \mathrm{pred}(b)}F_pP_{p,b}
\end{align}

The only exception is for the entry block of the program, which is always executed once to start the program. Therefore, we find

\begin{align}
F_r = 1 + \sum_{p \in \mathrm{pred}(r)}F_pP_{p,r}
\end{align}

This formulation creates a system of linear equations that can be solved to find $F_b$ for each block $b$.
One situation that we have to discuss in more detail are infinite loops. For our purposes, an infinite loop is a connected component of the CFG that does not reach an exit block. For all cases, where the infinite loop can be reached from the entry block, the expected number of times each block in the infinite loop will be executed is infinite. Since we want to prioritize instructions that are not part of infinite loops, since the fidelity of any execution that enters an infinite loop will always be zero independent of how the qubits are allocated.

To circumvent this problem, we do not consider back edges to blocks that cannot reach an exit block when calculating expected executions. This change is sufficient to ensure the system of equations for execution frequencies always has a solution.

To efficiently allocate the program, we further have to determine the transition probabilities between basic blocks in the CFG. The transition probabilities are in general hard to determine and depend on the quantum program. Our implementation allows users to specify probabilities in pragmas inserted before conditional branch instructions. These pragmas allows users to observe empirical branch probabilities and use them to improve allocation quality. If a conditional branch does not have a branch probability pragma, our implementation uses 0.5 as the default for the branch probability.

%here

% \section{Allocation Consistency}
% \label{sec:swaps}

\subsection{SWAP Placement Protocol}

In this section, we will discuss how the consideration of quantum control flow changes the SWAP placement problem. On most currently existing NISQ devices, swap-placement is needed if the quantum device has a limited connectivity between qubits. Considering control flow, we also have to consider the possibilities of inconsistent allocations. For example, if a swap is inserted into the middle of block $b$ and $b$ can reach itself, then the first time $b$ is entered the swap will not yet have been applied and the second time $b$ is entered the swap will have been applied. Thus, the swap-insertion problem motivates the {\textit{Routing Invariant}}. The routing invariant states that there must be a unique, statistically known allocation at every point in the program.

To restore the routing invariant and maintain allocation consistency, an inverse swap must be inserted at some point after the original swap but before control returns to $b$. This problem is not unique to loops.
Consider a conditional branch where one arm contains a swap and the other arm does not. Then the qubit allocation at the end of the branch depends dynamically on which branch of the arm was executed. In this situation, an inverse swap would be necessary at the end of the conditional arm that contained the original swap to ensure a consistent allocation after the branch.

A naive solution to this problem would be to insert an inverse swap as soon as possible after every swap. In effect, swaps would be inserted as necessary and last for the duration of only one instruction. However, we will show that such eager inverse swap insertion is contrary to the lazy swap insertion strategy~\cite{finigan2018} and could lead to far more swaps than necessary.

% \subsection{Inverse SWAP Insertion Protocol}
% \label{sec:invswaps}

For a more reliable solution, we use a strategy from the compilers literature~\cite{aho2008}, in particular the \textit{dominator} and \textit{strict dominator} relations between basic blocks.
Basic block $b$ \emph{dominates} block $b^\prime$ if every path from the entry block to $b^\prime$ includes $b$. In other words, $b$ dominates $b^\prime$ if $b$ must be executed in order from $b^\prime$ to be executed. Basic block $b$ \emph{strictly dominates} block $b^\prime$ if $b$ dominates $b^\prime$ and additionally $b \ne b^\prime$.
There exist efficient algorithms to calculate the dominators of all blocks in a CFG, which our implementation does as a preprocessing step~\cite{cooper2006}.

The rule for determining where to insert inverse swaps is the following: for each swap $s$ in basic block $b$, insert an inverse swap $s^{-1}$ on every edge $(b_1, b_2)$ such that $b$ dominates $b_1$ and $b$ does not strictly dominate $b_2$. In practice, inserting a swap on an edge $(b_1, b_2)$ means that our optimizer generates a new ``trampoline'' basic block $b^\prime$ and replaces the jump to $b_2$ in $b_1$ with a jump to $b^\prime$. $b^\prime$ contains any necessary inverse swaps followed by an unconditional jump to $b_2$ using the \textsc{JUMP} instruction. Inserting the inverse swaps any earlier is unnecessary because every basic block on a path from $b$ to $b_1$ is dominated by $b$. That means that swap $s$ will definitely have been executed and therefore cannot cause any inconsistencies in each of these blocks. Inserting the inverse swaps any later is too late, since there must exist some path from the entry block to the beginning of $b_2$ that does not go through $b$. This path does not contain the swap $s$, so it is possible to reach the beginning of $b_2$ with $s$ not in effect. This means that swap $s$ must be undone before reaching the beginning of $b_2$ to prevent inconsistencies.

Additionally, to preserve functional equivalency of the allocated circuit, for each \texttt{MEASURE} instruction in the input program, we adjust the corresponding physical qubit index to maintain consistency with the allocation at that specific point in the program, if necessary.

In our implementation, all consideration of control flow is encapsulated in the calculations of the swap set and fidelity bound done at each step of the local search. The higher level qubit allocation algorithms are therefore agnostic to the presence or absence of control flow.

\section{Proof-of-Concept Simulations}
\label{sec:sims}

In this section we illustrate and verify our implementation of qubit allocation considering control flow. The two hypotheses we test in these simulations are a correct basic block weighting and the consistency of our allocation methods. We have designed proof-of-concept statistical experiments using various instances of quantum simulations that test the components of these hypotheses and provide empirical evidence for the effectiveness our implementation laid out in this paper.

We begin by simulating the test circuit with control flow, previewed in Fig.~\ref{fig:control-flow}, on a fully-connected, noiseless quantum simulator using the Rigetti QVM~\cite{smith2017}. The set of operations in the test circuit consists solely of single-qubit rotational and two-qubit entanglement gates in the universal gate set native to Rigetti Quantum Processing Units (QPU), given by $\cal U$

\begin{equation}
\begin{split}
\mathcal{U} =&\bigcup_{q\in Q}\{R_x(k\pi, q): k\in\{-1,-\frac{1}{2},\frac{1}{2}, 1\}\cup\\              &\bigcup_{q\in Q}\{R_z(\phi, q): \phi\in[-\pi,\pi]\}\cup\\
            &\bigcup_{q\in Q}\{\textnormal{MEASURE}(q, c)\}\cup\\
            &\bigcup_{q_c,q_t\in E}\{CZ(q_c, q_t)\}
\end{split}
\end{equation}

where $Q$ is the set of physical qubits on the QPU, $E$ is the set of edges $(q_c, q_t)$ such that a controlled-$Z$ gate is permissible with control qubit $q_c$ and target qubit $q_t$, and MEASURE$(q, c)$ writes a measurement of qubit $q$ in the computational ($Z$) basis to classical bit $c$ ~\cite{smith2017}. The single-qubit Pauli rotations on qubit $q\in Q$ are given by:

$$R_x(\theta) =
\left[\begin{matrix}
\cos(\theta/2) & -i\sin(\theta / 2)\\
-i\sin(\theta/2) & \cos(\theta/2)
\end{matrix}\right]$$
$$R_z(\phi) =
\left[\begin{matrix}
\cos\left(\frac{\phi}{2}\right) -i \sin\left(\frac{\phi}{2}\right)& 0\\
0& \cos\left(\frac{\phi}{2}\right)+i\sin\left(\frac{\phi}{2}\right)
\end{matrix}\right]$$

\begin{figure}
      \includegraphics[width=0.45\textwidth]{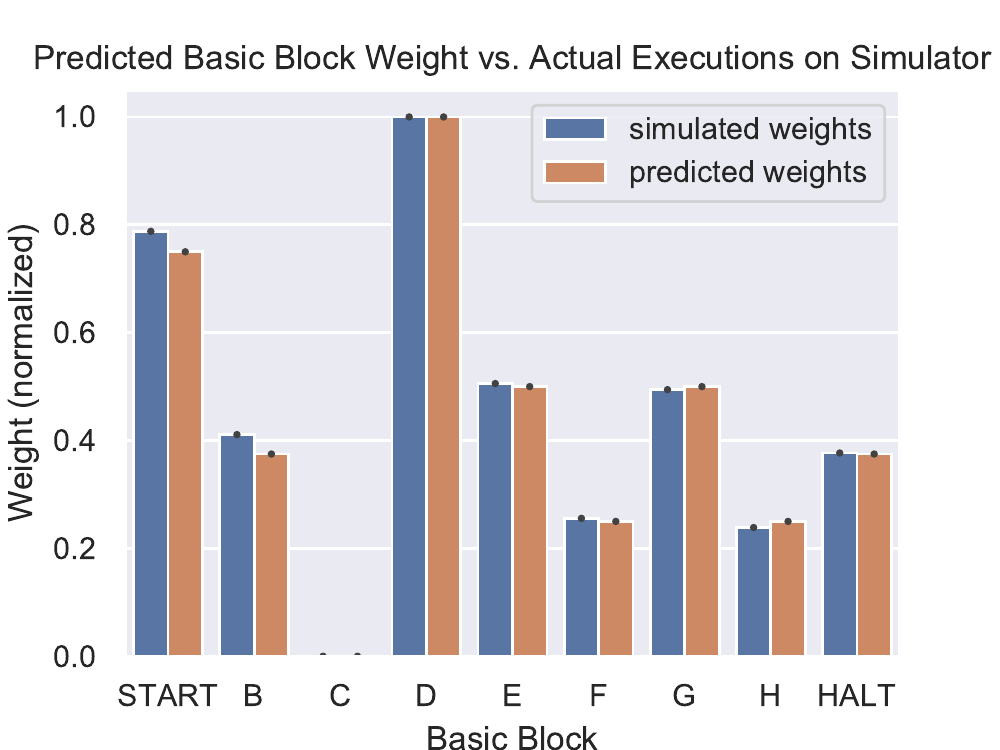}
      \caption{
      Comparison of predicted block weight and actual execution weight on simulator using the quantum algorithm outlined in Fig.~\ref{fig:control-flow}.}
      \label{fig:simulator}
 \end{figure}

In addition to the qubit operations, the test circuit consists of classical memory declaration, and control flow instructions: LABEL, JUMP, JUMP-WHEN, JUMP-UNLESS, and HALT such that the program exhibits the CFG structure shown in Fig.~\ref{fig:control-flow} left and is sufficiently complex to return a nontrivial distribution of measurement results over a given number of trials.
Prior to the simulation, we computed the predicted basic block weight of each block of the program using the algorithm presented above.
Next, we simulated the program 200 times independently and computed the normalized average number of executions of each block.
We found our normalized weights very closely matched the normalized mean number of executions, with an $R^2$ score of 0.96 and we show the results in Fig.~\ref{fig:simulator}.

In Fig.~\ref{fig:simulator}, we can see that basic block C was given a weight of $0$ and was never executed. This is due to the fact that there were no jump instructions leading to the block. Correspondingly our algorithm treated it with a weight of $0$. In summary, this experiment offers empirical evidence for the effectiveness of the block weight algorithm. We note that this accuracy is dependent on the actual branch probabilities. These branch probabilities are in general hard to determine and depend on the quantum program, i.e. full set of gates. For the program at hand, the use of $0.5$ as predicted branch probability performs very accurately with respect to the simulated basic block weights.

To test the overall performance of our qubit allocation scheme including control flow, we use a series of simulations. We test that, given an input program and a target QPU back-end, our output circuit allocated to the hardware is functionally equivalent (in practice) to the input program, as functional equivalence implies allocation consistency. More specifically, because the classical output and execution path of a quantum circuit with control flow is inherently probabilistic, we show that the distributions of output measurements of the original and allocated programs statistically have a high overlap for a sufficient number of trials.

\begin{figure}

  \includegraphics[width=\linewidth]{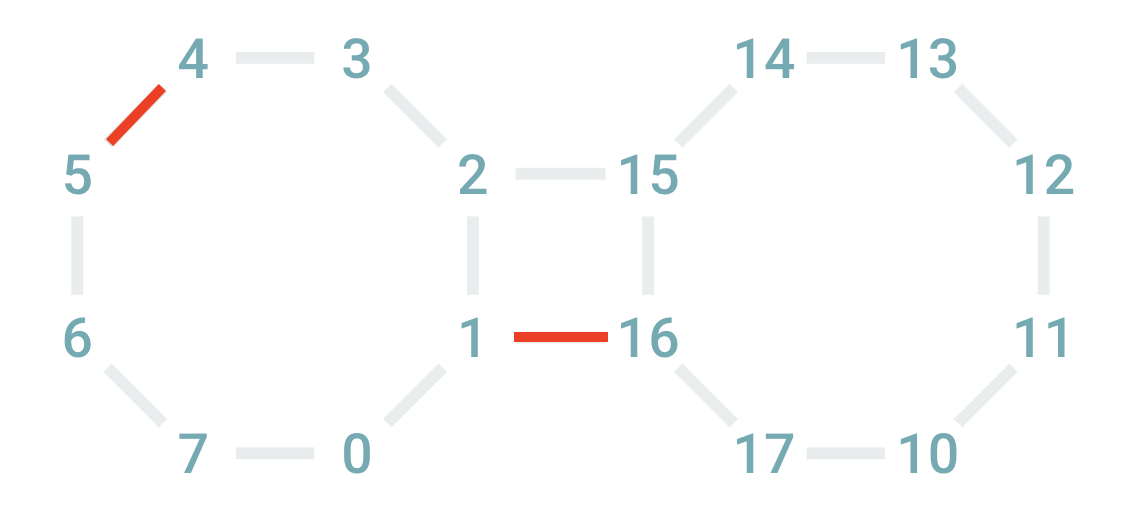}
  \caption{Qubit topology of Aspen-4-16Q-A at time of allocation/simulations. Two-qubit operations on edges (1, 16) and (4, 5) were prohibited.}
  \label{fig:aspen}
\end{figure}

We choose a test circuit, which operates on $15$ logical qubits, and measures $6$ of them at the end of the execution. Therefore, the dimensions of the qubit operation matrices are $2^{15}\times 2^{15}$ and the number of possible measurement results, denoted as $N$, is $2^6=64$.  We then choose the Rigetti Aspen-4-16Q-A device as our target hardware, as it has a sufficient number of physical qubits and its topology is sparse enough to make allocation nontrivial. We then ran our local search allocation algorithm on the same test circuit from above targeting the $16$-qubit Aspen back-end (topology described in ~\ref{fig:aspen}) under $2$ conditions: (1) CF-aware: In this condition, we allocate the circuit weighting for control flow and optimizing for reported gate fidelities of the device; (2) CF-unaware: Under this condition, we allocate the circuit optimizing for the gate fidelities of the device, but with no weighting for control flow (i.e. uniform basic block weights). Additionally, instead of using the inverse swap insertion implementation strategy, we instead use the naive solution of inserting an inverse swap as soon as possible after every swap, so we can compare our CF-aware implementation with a naive implementation of allocation consistency with no weighting for control flow.

\begin{figure}

  \includegraphics[width=.7\linewidth]{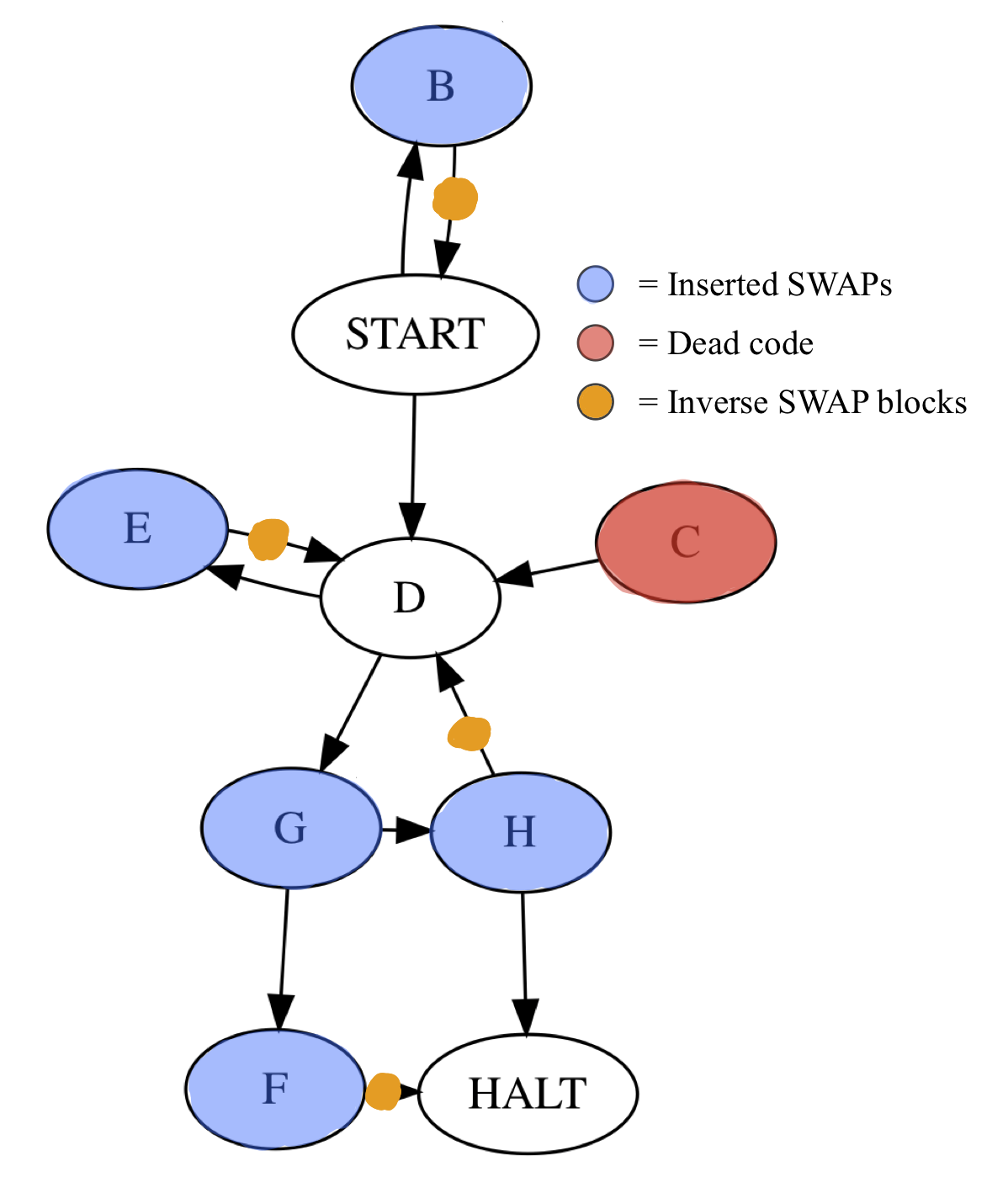}
  \caption{Control-flow graph of post-processed allocated circuit.}
  \label{fig:alloc}
\label{fig:test}
\end{figure}

For both allocation methods, the algorithm recognizes that Block C is unreachable and eliminates the dead code from the output circuit. Fig. ~\ref{fig:alloc} illustrates which basic blocks necessitates additional \texttt{SWAP} instructions for the allocations to adhere to the qubit topology of the Aspen device and also shows the locations of the inverse swap basic blocks from the CF-aware allocation method.


Before we can compare the performance of the two allocation methods on the Aspen-4-16Q-A noisy simulator, we first need to test that both of the allocated circuits are functionally equivalent to the input circuit in practice, and retrieve the ideal distribution of the input circuit from which we can compare metrics of statistical overlap across the two allocation methods. In order to do so, we first simulate the input circuit (on a fully-connected, noise-free simulator) so we can have a reference to the overall ideal distribution of the input circuit. Due to the indeterminacy of control flow in quantum circuits, we must discern how many simulations of the input circuit are necessary to capture the overall distribution across all paths. Using the empirical distribution as the ideal distribution from running $10$ simulations of the circuit would be inaccurate, as $10$ is clearly not a sufficient number of trials to capture the ideal distribution over all paths, whereas using the empirical distribution from running millions of trials is certainly more accurate with respect to the ideal distribution, but is clearly impractical. Our goal is to thus find a sufficient number of trials $n$ such that when we run parallel independent simulations of the input circuit (without noise) for $n$ trials each, their empirical distributions overlap to a very high degree on average.

In this paper, we refer as overlap to the metric of \textit{Squared Statistical Overlap}~\cite{sso} (SSO) that measures the statistical overlap between the measured and expected probabilities for all $N$ states, defined as

$$SSO = \left(\sum_{j=1}^N\sqrt{e_jm_j}\right)^2$$

where $N$ is the number of possible states, $e_j$ is the expected probability of measuring state $j$ and $m_j$ is the experimental probability of measuring state $j$.

\begin{figure}
    \includegraphics[width=0.46\textwidth]{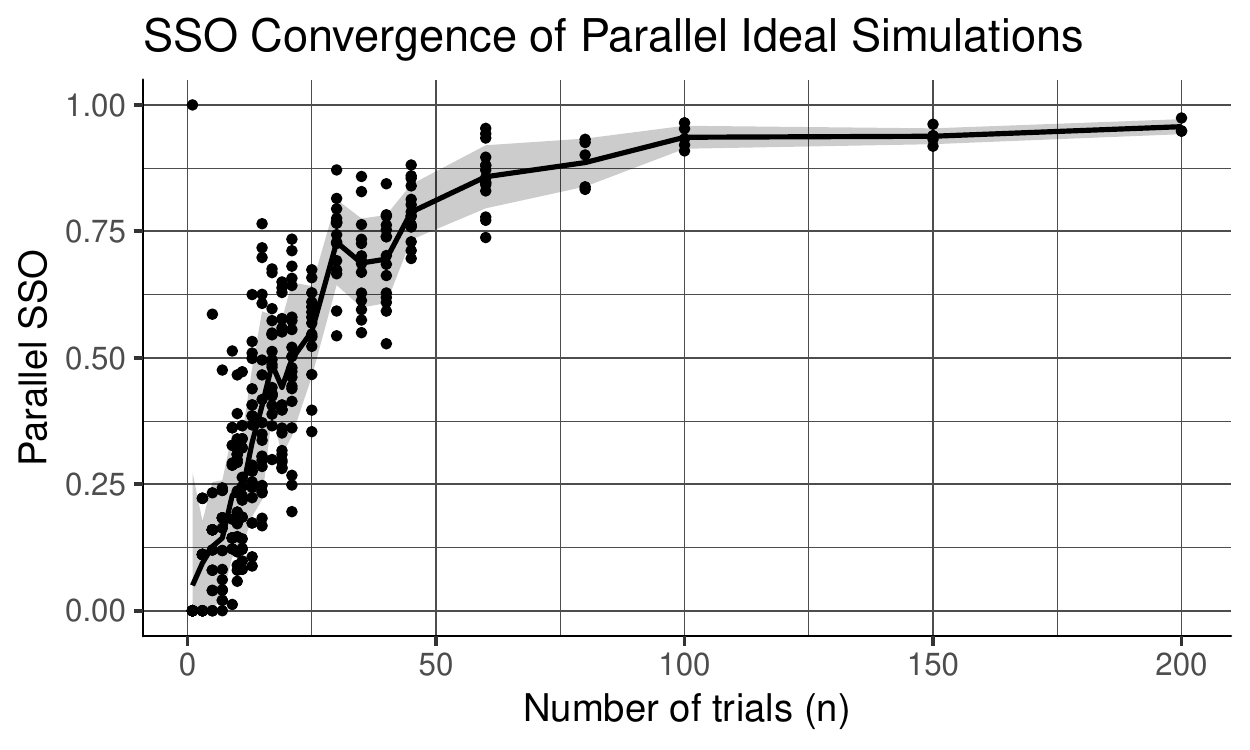}
    \caption{Each point represents the SSO of two independent simulations of the original test circuit for $n$ trials on a fully-connected and noise-free QVM. We include plotting the sample mean $\pm$ the sample standard deviation.}
    \label{fig:ideal_sso}
\end{figure}

\begin{figure}
    \includegraphics[width=0.48\textwidth]{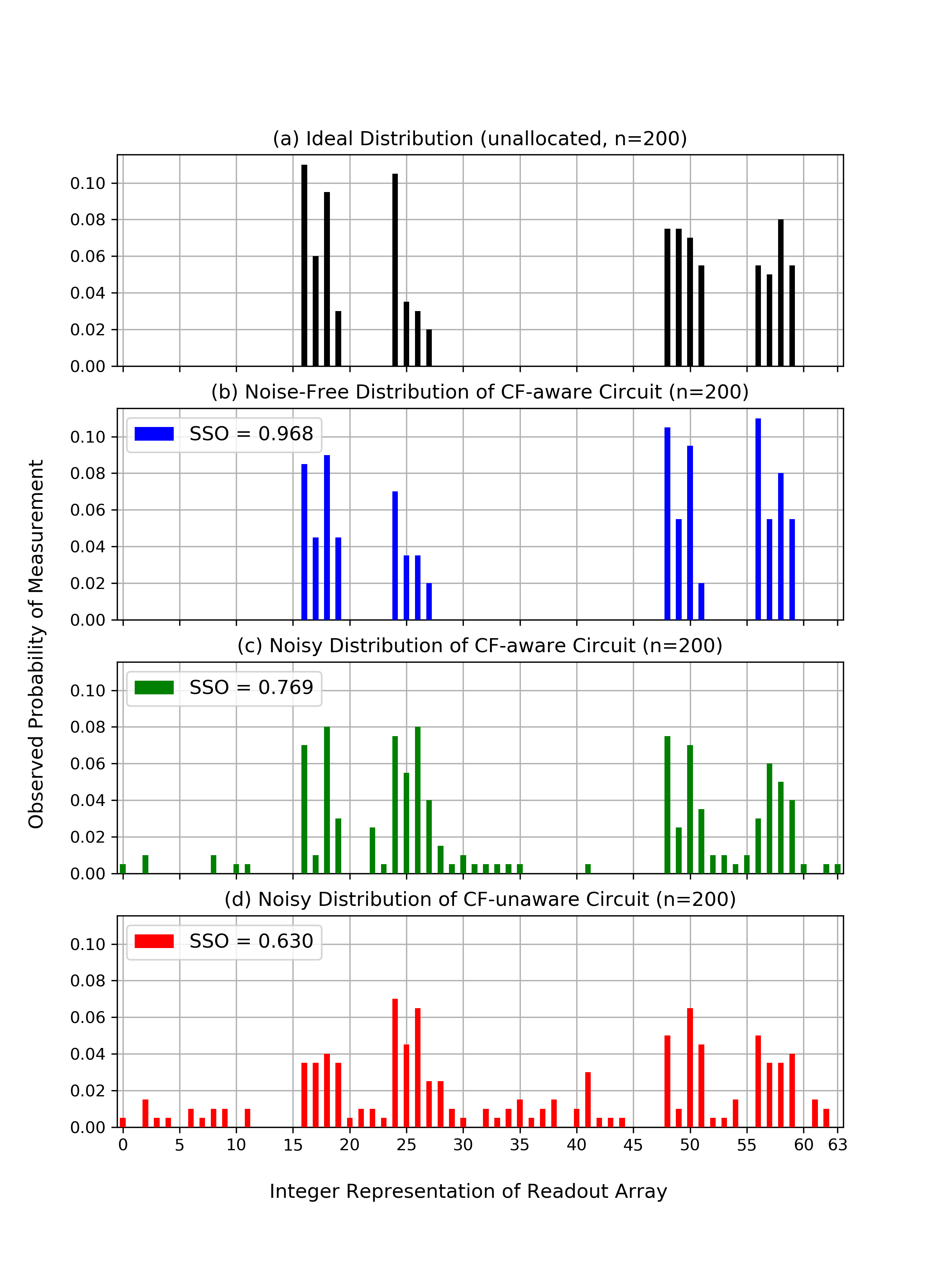}
    \caption{Visualization of observed measurement distributions for allocated and unallocated circuits. Experiment (a) ran on a fully-connected noise-free QVM, (b) on a noise-free QVM with the qubit topology of the Aspen device shown in Fig. ~\ref{fig:aspen}, and experiments (c) and (d) ran on a noisy QVM with the qubit topology and decoherence noise model of the Aspen-4-16Q-A QPU using the available Rigetti internal noise modeling functions that take into account gate execution times, $T_1$ and $T_2$ times, and readout fidelities for each physical qubit on the QPU.}
    \label{fig:dists}
\end{figure}

$SSO\in[0, 1]$, where $1$ signifies a perfect overlap of distributions and $0$ signifies no overlap. In Fig.~\ref{fig:ideal_sso} we see that the Parallel SSO asymptotically approaches perfect overlap (consistent with theory), and the data is heteroskedastic, as the variance of Parallel SSO decreases as $n$ increases. Because repeated simulations at $n=200$ trials have an average parallel SSO of $0.96$, we find that $n=200$ is a sufficient number of trials to capture the overall distribution of the input circuit. As a consequence, we uniformly sample one of the $n=200$ distributions and use it as our ideal distribution from which we can benchmark how the allocated circuits' overall distributions fare with respect to the ideal. The ideal distribution referenced in the remainder of the experiments are visualized in Fig.~\ref{fig:dists} (a).

\begin{figure}
    \includegraphics[width=.48\textwidth]{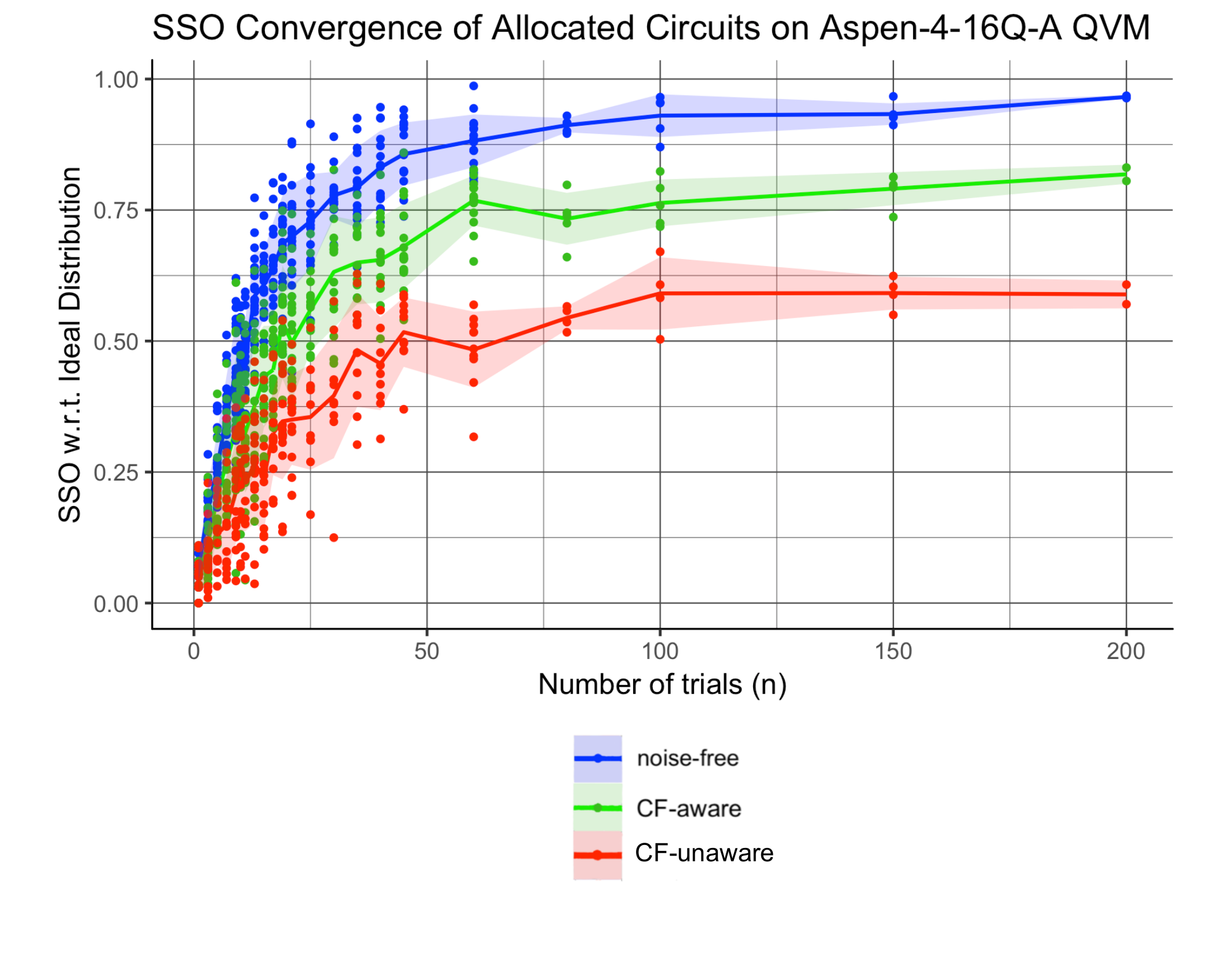}
    \caption{SSO Convergence of noise-free and noisy simulations of the two allocated circuits on Aspen$-4-16$Q$-$A QVM.}
    \label{fig:ssos}
\end{figure}

Next, to verify the claim of practical functional equivalence, we simulated our CF-aware allocated circuit on a noise-free QVM with the qubit topology of the Aspen device and plotted the SSO to the ideal distribution for repeated simulations of $n$ trials for $n$ ranging from $1$ to $200$. As shown in the plot in Fig.~\ref{fig:ssos}, the noise-free simulation shows the same rate of convergence and heteroskedasticity as that of Fig. ~\ref{fig:ideal_sso}, and results in a distribution matching the ideal distribution with an SSO of $0.968$ at $n=200$ trials, as seen in Fig. ~\ref{fig:dists} (b), which is consistent with the expected Parallel SSO for $n=200$ shown in Fig.~\ref{fig:ideal_sso}. The high SSO expresses that the probability that a functionally different circuit with inconsistent allocation would have a very similar rate of convergence and that high of a statistical overlap with the input circuit for $n=200$ trials is negligible. Therefore, these results confirm that the CF-aware allocated circuit is functionally equivalent in practice to the input circuit, and thus support our claim that our inverse swap insertion protocol maintains allocation consistency and adheres to the routing invariant.

When simulated on a noise-free QVM with the topology of the Aspen device, the allocated circuit from the CF-unaware allocation method with immediate inverse swap insertion showed almost identical results and is thus also functionally equivalent in practice to the input circuit. Now that we have verified allocation consistency and functional equivalence for the two allocated circuits to the input circuit, to compare the performance of the two methods, we run noisy simulations of the circuits on the Aspen QVM using a decoherence noise model instantiated from the topology and noise characteristics of the Aspen QPU at the time of allocation/simulation. Finally, we plot the progression of their SSOs to the ideal distribution for different values of $n$ in Fig. ~\ref{fig:ssos}. The resulting $n=200$ distributions from these noisy simulations of the CF-aware and CF-unaware allocated circuits are visualized in Fig. ~\ref{fig:dists} (c) and (d) with SSOs of $0.769$ and $0.630$, respectively. These results from our performance evaluation experiment on the noisy simulator show that on average, our implementation exhibits approximately a $14\%$ increase in SSO to the ideal simulation compared to the CF-unaware allocation method that does not take into account basic block weights or our inverse swap insertion protocol for $n=200$ trials, which has been shown to be a sufficient number of trials to capture the full distribution of this given test circuit. Though the CF-unaware method optimizes for the reported Aspen gate fidelities, it is our belief that its circuit fidelity for any robust noise model used in the CF-unaware allocation method will be dominated by its (in)ability to account for basic block weights and its inefficient inverse swap insertion strategy. We expect that this relative benefit of our implementation increases further with circuit complexity, however the time complexity of the local search, and any other exhaustive-search allocation algorithm, is not favorable to scaling on bigger and more complex circuits and devices, so compilers must employ probabilistic allocation techniques such as the hybrid algorithm in Ref.~\citenum{finigan2018}.

\subsection{Summary and Outlook}
\label{sec:summary}

In summary, we have developed a general method to efficiently incorporate programs with arbitrary control flow in the qubit allocation problem. Circuits with control flow rely on intermediate measurements, which no NISQ hardware currently offers, so we have validated and assessed our approach using various instances of the Rigetti QVM and its associated Quil instruction set architecture.

Our approach now introduces a powerful tool to quantum software and hardware developers alike, with the full flexibility of an on-chip, Turing-complete quantum programming language. We expect that the experiments described in this section will become adopted as a benchmarking standard for control-flow enabled quantum computation and simulation on a variety NISQ computing architectures.

Future research will test the performance of our implementation for a variety of control flow circuits across several devices/simulators and architectures and explore how the rate of convergence of the SSO in the experiments changes based on properties of the circuit/back-end, as this would allow us to potentially predict how many trials are sufficient to capture the overall distribution of the circuit. We also seek to address the question of how one may consider debugging a quantum program~\cite{huang2018, huang2019} that includes control flow commands, or how we can further reduce the noise of the computation by e.g. using randomized compilation approaches~\cite{wallman2016, ware2018} or other circuit optimizing routines. Another interesting route for improvement would be by using algorithms introduced for machine learning/data science application, such as a Bayesian optimizer~\cite{shahriari2016} considering the outcome of already performed quantum measurements to predict the weights of individual building blocks on the fly.

Our implementation introduces the first practical techniques to optimally allocate quantum programs onto NISQ devices, and will be seamlessly integrated with QPU execution when devices begin to support intermediate measurement of physical qubits. Enabling quantum control flow on real devices is a significant and essential milestone for the field of quantum computing, as state-of-the-art quantum algorithms, hybrid quantum-classical algorithms~\cite{peruzzo2014,mcClean2016}, and quantum error-correcting codes~\cite{kitaev1997, fowler2012} will benefit from control flow and Turing-complete instruction languages.

\subsection*{Acknowledgments}
\label{sec:ack}
This work is supported by Harvard University OTD's PSE Accelerator Grant. J.F. acknowledges fellowship support from the Deutsche Forschungsgemeinschaft (DFG) under Contract No. FL 997/1-1.  T.L. notes that this work is not associated with Google LLC or the Google Quantum AI Lab.

\bibliography{refs.bib}

\end{document}